\documentclass[prd,showpacs,preprintnumbers,amsmath,amssymb]{revtex4}
\usepackage{graphicx}% Include figure files
\usepackage{dcolumn}% Align table columns on decimal point
\usepackage{bm}% bold math

\begin{document}

\title{Effective Action of QED in Electric Field Backgrounds}
\author{Sang Pyo Kim}\email{sangkim@kunsan.ac.kr}
\affiliation{Department of Physics, Kunsan National University,
Kunsan 573-701, Korea} \affiliation{Asia
Pacific Center for Theoretical Physics, Pohang 790-784, Korea}

\author{Hyun Kyu Lee}\email{hyunkyu@hanyang.ac.kr}
\author{Yongsung Yoon}\email{cem@hanyang.ac.kr}
\affiliation{Department of Physics, Hanyang University, Seoul
133-791, Korea}

\date{\today}

\begin{abstract}
We use the evolution operator method to find the one-loop
effective action of scalar and spinor QED in electric field
backgrounds in terms of the Bogoliubov coefficient between the
ingoing and the outgoing vacua. We obtain the exact
one-loop effective action for a Sauter-type electric field, $E_0 \,{\rm
sech}^2 (t/\tau)$, and show that the imaginary part correctly yields
the vacuum persistence. The renormalized effective action shows the
general relation between the vacuum persistence and the total mean number
of created pairs for the constant and the Sauter-type electric field.
\end{abstract}

\maketitle

\section{Introduction}

More than seven decades ago, Sauter, Heisenberg and Euler, and
Weisskopf studied the effective action of a charged particle in a
constant electromagnetic field \cite{Sauter}. Using the proper time
method, Schwinger systematically derived the effective action in a
gauge invariant form in a constant electromagnetic field
\cite{Schwinger}. There have been since then many attempts to find
the effective actions in various configurations of electromagnetic
fields (for a review, see Ref. \cite{Dunne}).  The recent revival of
the effective action in strong electromagnetic fields is partly due
to a direct test of the strong QED in terrestrial experiments in the
near future \cite{Ringwald} and partly due to applications to
astrophysical objects with strong electromagnetic fields beyond the
critical strength.

In contrast to constant electromagnetic fields, finding the
effective action in inhomogeneous electromagnetic fields is a
non-trivial task. There have been recent attempts to find the
pair-production rate in inhomogeneous electric fields, in
particular, in localized electric fields either in space or time
\cite{Kim-Page02,Kim-Page06,Kim-Page07,GK,Dunne-Schubert,DWGS,KRX}.
A key idea is the electric-magnetic duality of the QED effective
action \cite{BYP}, according to which the effective action for a
constant magnetic field can be analytically continued to that for a
constant electric field. Though a Landau level in a constant
magnetic field is dual to the tunneling motion in a constant
electric field \cite{BPS}, in the real spacetime a particle
accelerates by an electric field, whose unbounded motion makes
non-trivial the task of finding the effective action from the
time-dependent state. Dunne and Hall used the resolvent technique
and directly found the effective action in time-dependent electric
fields \cite{Dunne-Hall} and Fried and Woodard found the effective
action for electric fields that depend on the light-cone time
coordinate \cite{Fried-Woodard}.

Recently, two of us (SKP and HKL) applied the evolution operator
method to find the effective action of scalar QED
in a constant electric field
\cite{Kim-Lee08}. In this method the ingoing vacuum evolves to the
outgoing vacuum via the evolution operator. The evolution operator
is given by a phase part and a squeeze operator, whose parameters
are determined by Bogoliubov coefficients. The effective action is
then defined through the scattering amplitude between the ingoing
and the outgoing vacua
\cite{Schwinger2,Nikishov1,DeWitt,AHN,Nikishov2,FGS,GGT}. The
renormalized effective action for a constant electric field, given by a
Bogoliubov coefficient for each momentum, correctly yields not only
the vacuum polarization but also the vacuum persistence
\cite{Kim-Lee08,Nikishov1,AHN,Nikishov2,FGS,GGT}. However, it is still an open
question whether the renormalized effective action for a time-dependent
electric field or an inhomogeneous electric field can satisfy
the general relation between the vacuum persistence and the total mean number of
created pairs.

The main purpose of this paper is to further develop the evolution
operator method in scalar QED, first by clarifying the
renormalization procedure in a constant electric field and then by
applying it to time-dependent electric fields. In particular, we
find the exact one-loop effective action for the Sauter-type
electric field, $E_0 \,{\rm sech}^2 (t/\tau)$, together with or
without a constant magnetic field. Finally, we extend the evolution
operator method to spinor QED and find the exact one-loop effective
action in the Sauter-type electric field. It is found that our
effective action in spinor QED under the Sauter-type electric field
shows the correct limiting behaviors, in comparison with the
effective action obtained by the resolvent technique
\cite{Dunne-Hall}. It is further shown that the imaginary parts of
the effective action in scalar and spinor QED satisfy the general
relations between the vacuum persistence and the total mean number of
created scalar particles and fermions.

The organization of this paper is as follows. In Sec. II, we
elaborate the evolution operator method for scalar QED by clarifying
the renormalization procedure and then extend it to spinor QED. This
is done by expressing the Bogoliubov transformation in terms of the
evolution operator and then the evolution operator by a two-mode
squeeze operator. We find the mean number of created pairs and the
vacuum persistence in terms of squeeze parameters, which are, in
turn, determined by the Bogoliubov coefficients, and finally obtain
the effective action in scalar and spinor QED. In Sec. III, we apply
the method to a constant electric field and clarify the
renormalization procedure in scalar and spinor QED. In Sec. IV, we
find the exact one-loop effective action for a Sauter-type electric
field together with or without a constant magnetic field in scalar
and spinor QED, which is the main result of this paper.

\section{Effective Action in Electric Fields}

The effective action of a charged particle has been studied in
various configurations of magnetic fields. The charged particle in a
magnetic field has a discrete spectrum from bounded motions, leading
to the effective action. However, in a strong electric field, a
virtual pair from the Dirac sea gains a sufficient potential energy
to separate over the Compton wavelength and to materialize as a real
pair. In the time-dependent gauge, the electric field provides a
potential barrier, over which the charged particle scatters from an
ingoing state to an outgoing one. To find the effective action in
electric fields is not straightforward because unbounded motions
cause instability of the vacuum. In this section we shall introduce
a new method to calculate the effective action at zero temperature
using the evolution operator, which can be factored into a phase
factor and a squeeze operator.

In the time-dependent gauge, a time-dependent electric field along
the $z$-direction is given by $E(t) = - \partial A_z (t)/\partial
t$. The Fourier component of the Klein-Gordon equation for scalar
QED and the spin diagonal component of the Dirac equation for spinor
QED satisfy [in units with $\hbar = c = 1$ and with metric signature
$(+, -, -, -)$]
\begin{eqnarray}
\Bigl[\partial_t^2 + m^2 + {\bf k}_{\perp}^2 + (k_z + q A_z (t))^2 +
2 i \sigma q E(t) \Bigr] \phi_{{\bf k} \sigma} (t) = 0,
\label{time-comp}
\end{eqnarray}
where $\sigma = 0$ for scalar particles and $\sigma = \pm 1/2$ for
spin-1/2 fermions.

\subsection{Scalar QED}

\subsubsection{Evolution Operator in terms of Squeeze Operator}

The ingoing vacuum is annihilated by $a_{\bf k} (t_{\rm in} = -
\infty)$ for particle and $b_{\bf k} (t_{\rm in} = - \infty)$ for
antiparticle for each momentum, and the outgoing vacuum is similarly
defined by $a_{\bf k} (t_{\rm out} = \infty)$ and $b_{\bf k} (t_{\rm
out} = \infty)$. These operators are related through the Bogoliubov
transformation \cite{Kim-Lee07}
\begin{eqnarray}
a_{{\bf k}, {\rm in}} &=& \mu^*_{\bf k} a_{{\bf k}, {\rm out}} -
\nu^*_{\bf k} b^{\dagger}_{{\bf k}, {\rm out}}, \nonumber\\
b_{{\bf k}, {\rm in}} &=& \mu^*_{\bf k} b_{{\bf k}, {\rm out}} -
\nu^*_{\bf k} a^{\dagger}_{{\bf k}, {\rm out}}, \label{in-out}
\end{eqnarray}
where
\begin{eqnarray}
\mu_{\bf k} &=& i \Bigl(\phi^*_{\bf k} (t_{\rm out})
\dot{\phi}_{\bf k} (t_{\rm in}) - \dot{\phi}^*_{\bf k} (t_{\rm
out})
\phi_{\bf k} (t_{\rm in})\Bigr), \nonumber\\
\nu_{\bf k} &=& i \Bigl(\phi^*_{\bf k} (t_{\rm out})
\dot{\phi}^{*}_{\bf k} (t_{\rm in}) - \dot{\phi}^*_{\bf k}
(t_{\rm out}) \phi^{*}_{\bf k} (t_{\rm in}) \Bigr).
\label{sol-bcoef}
\end{eqnarray}
with the wronskian condition, $\dot{\phi}^*_{\bf k} (t) \phi_{\bf k} (t)
- \dot{\phi}_{\bf k} (t) \phi^*_{\bf k} (t) = i$.
These coefficients satisfy the relation
\begin{eqnarray}
|\mu_{\bf k}|^2 - |\nu_{\bf k}|^2 = 1.
\end{eqnarray}
The inverse Bogoliubov transformation is
\begin{eqnarray}
a_{{\bf k}, {\rm out}} &=& \mu_{\bf k} a_{{\bf k}, {\rm in}} +
\nu^*_{\bf k} b^{\dagger}_{{\bf k}, {\rm in}}, \nonumber\\
b_{{\bf k}, {\rm out}} &=& \mu_{\bf k} b_{{\bf k}, {\rm in}} +
\nu^*_{\bf k} a^{\dagger}_{{\bf k}, {\rm in}}. \label{out-in}
\end{eqnarray}
The Bogoliubov coefficients $\mu_{\bf k}$ and $\nu_{\bf k}$
implicitly depend on the gauge potential $A_z$ through Eq.
(\ref{time-comp}), so they will be denoted by $\mu_{\bf k} (A)$ and
$\nu_{\bf k} (A)$. For a constant electric field in Sec. III and a
Sauter-type electric field in Sec. IV, the coefficients directly
found from the exact solution are free of the dynamical phases of
evolution. However, a caveat is that the dynamical phase of Eq.
(\ref{sol-bcoef}) should be removed in the effective action to
account only interactions with the electric field background. For
instance, even a free scalar field for $A_z = 0$ with the solution
$\varphi_{\bf k} = e^{- i \omega_{\bf k} t}/\sqrt{2 \omega_{\bf
k}}$, $(\omega_{\bf k} = \sqrt{{\bf k}^2 + m^2})$, has a dynamical
phase given by
\begin{eqnarray}
\mu_{\bf k} (A = 0) = e^{i \int_{t_{\rm in}}^{t_{\rm out}}
\omega_{\bf k} dt},
\end{eqnarray}
and $\nu_{\bf k} = 0$. Thus, the Bogoliubov coefficients in Eq.
(\ref{sol-bcoef}) carry the information of both the dynamical phase
and the gauge potential.

To express the outgoing vacuum as multi-particle states of the
ingoing vacuum, we write the Bogoliubov transformation
(\ref{out-in}) as a unitary transformation
\begin{eqnarray}
a_{{\bf k}, {\rm out}} = U_{\bf k} a_{{\bf k}, {\rm in}}
U^{\dagger}_{\bf
k}, \nonumber\\
b_{{\bf k}, {\rm out}} = U_{\bf k} b_{{\bf k}, {\rm in}}
U^{\dagger}_{\bf k}. \label{sq op}
\end{eqnarray}
Here, the evolution operator,
\begin{eqnarray}
U_{\bf k} (A) =  S_{\bf k} (A) P_{\bf k} (A),
\end{eqnarray}
is factored by the overall phase factor and the two-mode squeeze
operator \cite{Caves-Schumaker1,Caves-Schumaker2}
\begin{eqnarray}
P_{\bf k} &=&  \exp \Bigl[i \theta_{\bf k} \Bigl(a^{\dagger}_{{\bf
k}, {\rm in}}a_{{\bf k}, {\rm in}} +
b^{\dagger}_{{\bf k}, {\rm in}} b_{{\bf k}, {\rm in}} + 1 \Bigr) \Bigr], \nonumber\\
S_{\bf k} &=& \exp \Bigl[ r_{\bf k} \Bigl( a_{{\bf k}, {\rm in}}
b_{{\bf k}, {\rm in}} e^{- 2i \vartheta_{\bf k}} - a^{\dagger}_{{\bf
k}, {\rm in}} b^{\dagger}_{{\bf k}, {\rm in}} e^{2i \vartheta_{\bf
k}} \Bigr) \Bigr],
\end{eqnarray}
where the squeeze parameter $r_{\bf k}$, the squeeze angle
$\vartheta_{\bf k}$, and the overall phase angle $\theta_{\bf k}$
are determined by
\begin{eqnarray}
\mu_{\bf k} &=& e^{- i \theta_{\bf k}} \cosh r_{\bf k}, \nonumber\\
\nu_{\bf k}^* &=& - e^{-i \theta_{\bf k}} (e^{2 i \vartheta_{\bf k}}
\sinh r_{\bf k} ). \label{sq pa}
\end{eqnarray}
As the outgoing vacuum is the two-mode squeezed vacuum
\begin{eqnarray}
\vert 0, {\rm out} \rangle = \prod_{\bf k} U_{\bf k} (A) \vert 0,
{\rm in} \rangle = U(A) \vert 0, {\rm in} \rangle,
\end{eqnarray}
the scattering amplitude is given by
\begin{eqnarray}
\langle 0, {\rm out} \vert 0, {\rm in} \rangle = \langle 0, {\rm in}
\vert U^{\dagger} \vert 0, {\rm in} \rangle = \prod_{\bf k} e^{ i
\theta_{\bf k}} \langle 0, {\rm in} \vert S^{\dagger}_{\bf k} \vert
0, {\rm in} \rangle.
\end{eqnarray}
The squeeze operator can further be factored as
\cite{Caves-Schumaker2}
\begin{eqnarray}
S_{\bf k} &=& \exp \Bigl[\xi_{\bf k} a^{\dagger}_{{\bf k}, {\rm in}}
b^{\dagger}_{{\bf k}, {\rm in}} \Bigr]  \exp \Bigl[
\frac{\gamma_{\bf k}}{2} \Bigl( a^{\dagger}_{{\bf k}, {\rm in}}
a_{{\bf k}, {\rm in}} + b^{\dagger}_{{\bf k}, {\rm in}} b_{{\bf k},
{\rm in}}+1 \Bigr) \Bigr] \exp \Bigl[ - \xi_{\bf k}^* a_{{\bf k},
{\rm in}} b_{{\bf k}, {\rm in}} \Bigr],
\end{eqnarray}
where
\begin{eqnarray}
\xi_{\bf k} &=& - e^{2 i \vartheta_{\bf k}} \tanh r_{\bf k},
\nonumber\\
\gamma_{\bf k} &=& \ln (1 - |\xi_{\bf k}|^2 ) = - 2 \ln (\cosh
r_{\bf k} ).
\end{eqnarray}

\subsubsection{Mean Number of Pairs and Vacuum Persistence}

Using the squeeze operator, we now find the vacuum persistence and
the mean number of created pairs. Due to charge neutrality, the
multi-particle state for $n$-pairs consists of equal number of
particles and antiparticles, which is compactly denoted as $\vert
n_{\bf k}, t \rangle = (a_{\bf k}^{\dagger n} (t) /\sqrt{n!})
(b_{\bf k}^{\dagger n} (t) /\sqrt{n!})\vert 0; t \rangle$. The
probability for $n$-pairs with momentum ${\bf k}$ to be created from
the vacuum is
\begin{eqnarray}
P_n ({\bf k}) = | \langle n_{\bf k}, {\rm out} \vert 0, {\rm in}
\rangle |^2  = | \langle n_{\bf k}, {\rm in} \vert S^{\dagger}_{\bf
k} \vert 0, {\rm in} \rangle |^2 = e^{\gamma_{\bf k}} |\xi_{\bf
k}|^{2n}.
\end{eqnarray}
Note that $P_0 = e^{\gamma_{\bf k}}$ and $P_1 = e^{\gamma_{\bf k}}
|\xi_{\bf k}|^{2}$ so that $P_n = P_0 (P_1/P_0)^n$ and $\sum_{n =
0}^{\infty} P_n = 1$ for each ${\bf k}$. Thus, the mean number of
pairs created from the vacuum for each momentum per unit volume is
\begin{eqnarray}
{\cal N}_{\bf k} = \sum_{n = 0}^{\infty} n P_n ({\bf k}) = \sinh^2
r_{\bf k} = |\nu_{\bf k}|^2.
\end{eqnarray}
The vacuum persistence is
\begin{eqnarray}
| \langle 0, {\rm out} \vert 0, {\rm in} \rangle |^2 &=& \prod_{\bf
k} P_0 ({\bf k})  =  \prod_{\bf k} \frac{1}{\cosh^2 r_{\bf k}} =
\prod_{\bf k} \frac{1}{|\mu_{\bf k}|^2}.
\end{eqnarray}
Alternatively, the vacuum persistence is the probability for the
ingoing vacuum to remain the outgoing vacuum
\begin{eqnarray}
| \langle 0, {\rm out} \vert 0, {\rm in} \rangle |^2 = \prod_{\bf k}
\Bigl[ 1 - \sum_{n = 1}^{\infty} P_n ({\bf k}) \Bigr] = \prod_{\bf
k} \frac{1}{\cosh^2 r_{\bf k}}.
\end{eqnarray}

\subsubsection{Effective Action}

Following Refs.
\cite{Kim-Lee08,Schwinger2,Nikishov1,DeWitt,AHN,Nikishov2,FGS,GGT},
the effective action at zero temperature is defined by the
scattering amplitude as
\begin{eqnarray}
e^{i S^{\rm sc}_{\rm eff}} = e^{i \int dt d^3{\bf x} {\cal L}^{\rm
sc}_{\rm eff}} = \langle 0, {\rm out} \vert 0, {\rm in} \rangle.
\end{eqnarray}
In fact, the effective action is equivalent to the usual form
\begin{eqnarray}
e^{i S^{\rm sc}_{\rm eff}} = \int {\cal D} \phi^* {\cal D} \phi e^{i
\int {\cal L}^{\rm sc} (A)}.
\end{eqnarray}
After a gymnastic of algebra, the effective action is given by
\begin{eqnarray}
\langle 0, {\rm out} \vert 0, {\rm in} \rangle = \prod_{\bf k}
\frac{1}{\mu^*_{\bf k}}. \label{0-tem}
\end{eqnarray}
The effective action may be found directly from the ingoing and the
outgoing vacuum wave functional, for instance, from the ground state
for $a_{\bf k}$ and $b_{\bf k}$,
\begin{eqnarray}
\langle 0_{\bf k}, {\rm out} \vert 0_{\bf k}, {\rm in} \rangle =
\Biggl[ \frac{\varphi^*_{\bf k} (t_{\rm in})  \varphi_{\bf k}
(t_{\rm out})}{|\varphi^*_{\bf k} (t_{\rm in})  \varphi_{\bf k}
(t_{\rm out})|} \Biggr] \frac{1}{\mu^*_{\bf k}}. \label{wave-eff}
\end{eqnarray}
Here, the dynamical phase in the square bracket, which is, for instance,
$e^{- i \int_{t_{\rm in}}^{t_{\rm out}} \omega_{\bf k} dt}$ for
$A=0$, is canceled by that of $\mu^*_{\bf k}$, so the effective
action (\ref{wave-eff}) is independent of the dynamical phase, as in
Eq. $(\ref{0-tem})$.

Finally, we find the effective action per unit
volume
\begin{eqnarray}
{\cal L}^{\rm sc}_{\rm eff} = i \sum_{\bf k} \ln (\mu^*_{\bf k}),
\label{eff-E}
\end{eqnarray}
where the summation is over all possible states in the momentum
space. It can be shown that the
vacuum persistence
\begin{eqnarray}
|\langle 0, {\rm out} \vert 0, {\rm in} \rangle |^2 = e^{ - 2 ({\rm
Im} {\cal S}^{\rm sc}_{\rm eff})} = e^{ - V \sum_{\bf k} \ln (1+
{\cal N}^{\rm sc}_{\bf k})},
\end{eqnarray}
where $V$ is the volume, leads to the exact relation between the
imaginary part and the total mean number of created scalar pairs
\begin{eqnarray}
2 ({\rm Im} {\cal L}^{\rm sc}_{\rm eff}) = \sum_{\bf k} \ln (1+
{\cal N}^{\rm sc}_{\bf k}). \label{scalar-rel}
\end{eqnarray}
The relation (\ref{scalar-rel}) is generally true for any electric
field background. In the weak-field limit, $(|\nu_{\bf k}|^2 \ll
1)$, twice of the imaginary part of the effective action per unit
volume is the total mean number of created pairs, $2 ({\rm Im} {\cal
L}^{\rm sc}_{\rm eff}) \approx \sum_{\bf k} {\cal N}^{\rm sc}_{\bf k} = {\cal
N}^{\rm sc}$.

However, it should be pointed out that the effective action (\ref{eff-E})
and thereby the general relation (\ref{scalar-rel}) involve diverging terms,
which require some proper regularization scheme to yield
renormalized ones. In this paper we shall show
that the renormalized effective action in scalar QED\emph{} indeed satisfies the general
relation (\ref{scalar-rel}) for a constant and a Sauter-type time-dependent
electric field.

\subsection{Spinor QED}

In spinor QED, the Bogoliubov transformation between the ingoing and
the outgoing particle and antiparticle operators, $b_{n}, d_{n}$, is
given by
\begin{eqnarray}
b_{n,{\rm out}} &=& \mu_{n} b_{n,{\rm in}} + i\nu_{n}^{*}d_{n,{\rm
in}}^{\dagger}, \nonumber \\ d_{n,{\rm out}} &=& \mu_{n} d_{n,{\rm
in}} - i\nu_{n}^{*}b_{n,{\rm in}}^{\dagger}, \label{spinor-bogol}
\end{eqnarray}
where $n = ({\bf k}, \sigma)$ denotes the momentum and spin states,
$\sigma = \pm 1/2$. The Bogoliubov coefficients satisfy the relation
\begin{eqnarray}
|\mu_{n}|^{2}+|\nu_{n}|^{2}=1.
\end{eqnarray}
The Bogoliubov transformation can be written as a unitary
transformation \cite{Fan}
\begin{eqnarray}
b_{n,{\rm out}} &=& U_{n}b_{n,{\rm in}}U_{n}^{\dagger}, \nonumber
\\ d_{n,{\rm out}} &=& U_{n}d_{n,{\rm in}}U_{n}^{\dagger},
\label{spinor-unitary}
\end{eqnarray}
where the evolution operator is factored into the overall phase
factor and the two-mode squeeze operator for fermions as
\begin{equation}
U_{n}=e^{\xi_{n}b_{n,{\rm in}}^{\dagger}d_{n,{\rm in}}^{\dagger}}
e^{(\frac{\gamma_{n}}{2}+i\theta_{n})(b_{n,{\rm
in}}^{\dagger}b_{n,{\rm in}}+d_{n,{\rm in}}^{\dagger}d_{n,{\rm
in}}-1)} e^{e^{2i\theta_{n}}\xi_{n}^{*}b_{n,{\rm in}}d_{n,{\rm
in}}}. \label{spinor-squeeze}
\end{equation}
The Bogoliubov coefficients are determined by three real parameters
$\theta_{n}, \vartheta_{n}, r_{n}$ for the evolution operator as
follows
\begin{eqnarray}
\mu_{n} &=& e^{-i \theta_{n}}\cos r_{n}, \nonumber \\ \nu_{n}^{*} &
=& - e^{-i\theta_{n}} (e^{2i\vartheta_{n}} \sin r_{n}), \nonumber
\\ \gamma_{n} &=& -2\ln(\cos r_{n}), \nonumber \\ \xi_{n} &=&
ie^{2i \vartheta_{n}} \tan r_{n}. \label{spinor-parameters}
\end{eqnarray}

The effective action defined as the scattering amplitude,
\begin{eqnarray}
e^{iS^{\rm sp}_{\rm eff}} = \langle 0,{\rm in} \vert \prod_{n}
U_{n}^{\dagger} \vert 0,{\rm in} \rangle,
\end{eqnarray}
leads, with the aid of Eq. (\ref{spinor-squeeze}), to the effective action per
unit volume for spinor QED in the form
\begin{eqnarray}
{\cal L}^{\rm sp}_{\rm eff} = - i \sum_{n} \ln (\mu^{*}_{n}),
\label{spinor-E}
\end{eqnarray}
where the summation $n$ is over all possible momentum ${\bf k}$ and
spin $\sigma$. The mean number of pairs created from the vacuum for
each state $n$ is calculated as
\begin{eqnarray}
{\cal N}^{\rm sp}_{n} = |\langle 1_{n},{\rm out}\vert 0, {\rm in}
\rangle|^{2}=\sin^{2} r_{n} = 1 - |\mu_{n}|^{2}.
\end{eqnarray}
Therefore, the vacuum persistence
\begin{eqnarray}
|\langle 0, {\rm out} \vert 0, {\rm in} \rangle |^2 = e^{ - 2 ({\rm
Im} {\cal S}^{\rm sp}_{\rm eff})} = e^{ V \sum_{n} \ln (1- {\cal
N}^{\rm sp}_{n})},
\end{eqnarray}
where $V$ is the volume, leads to the exact relation between the
imaginary part and the total mean number of created pairs
\begin{eqnarray}
2 ({\rm Im} {\cal L}^{\rm sp}_{\rm eff}) = - \sum_{n} \ln (1 - {\cal
N}^{\rm sp}_{n}). \label{spin-rel}
\end{eqnarray}
The relation (\ref{spin-rel}) is generally true for any electric
field background in spinor QED.

As in the case of scalar QED, the spinor effective action (\ref{spinor-E})
and the general relation (\ref{spin-rel}) are not renormalized
yet. In the next sections we shall put forth
a regularization scheme, obtain the
renormalized effective action and show the general relation (\ref{spin-rel}).

\section{Effective Action in a Constant Electric Field}

As the first application of the method in Sec. II, we find the
effective action for a constant electric field. In the
time-dependent gauge $A_z = - E t$, the Fourier-component of the
Klein-Gordon or the spin diagonal Dirac equation takes the form
\begin{eqnarray}
\Bigl[\partial_t^2 + m^2 + {\bf k}_{\perp}^2 + (k_z - qE t)^2 + 2 i
\sigma qE \Bigr] \phi_{\omega, {\bf k}} (t) = 0,
\end{eqnarray}
where $\sigma = 0$ for scalar QED and $\sigma = \pm 1/2$ for spinor
QED. The equation describes the scattering problem over an inverted
harmonic potential. The asymptotic ingoing and the outgoing vacuum
may be defined with respect to the asymptotically positive frequency
at $t = - \infty$ and $+ \infty$, respectively. A positive frequency
solution at $t = - \infty$ is, in general, scattered into a branch
of positive frequency and another branch of negative frequency at $t
= \infty$. The solution of the parabolic cylinder function
\begin{eqnarray}
\phi_{\omega, {\bf k}} (t) = D_p(z),
\end{eqnarray}
with
\begin{eqnarray}
z = \sqrt{\frac{2}{qE}} e^{i \pi/4} (k_z - qE t), \quad p = -
\frac{1}{2} - i \frac{m^2 + {\bf k}_{\perp}^2 + 2 i \sigma qE}{2
(qE)},
\end{eqnarray}
has an appropriate ingoing flux at $t = - \infty$ with respect to
the asymptotic form $D_p (z) \approx e^{- z^2/4} z^p$ for $|z| \gg
1$. In the other asymptotic region at $t = \infty$, the solution is
analytically continued to the form \cite{GR2}
\begin{eqnarray}
D_p (z) = e^{-i p \pi} D_{p} (-z) + \frac{\sqrt{2 \pi}}{\Gamma(-p)}
e^{-i (p+1)\pi/2} D_{- p-1} ( iz). \label{con E-sol}
\end{eqnarray}
Thus, we find the Bogoliubov coefficients
\begin{eqnarray}
\mu_{\bf k} = \frac{\sqrt{2 \pi}}{\Gamma(-p)} e^{- i (p+1)\pi/2},
\quad \nu_{\bf k} = e^{ -i p \pi}. \label{E-bcoef}
\end{eqnarray}

\subsection{Scalar QED}

From Eq. (\ref{eff-E}) for scalar QED, the effective action per unit
volume is given by
\begin{eqnarray}
{\cal L}^{\rm sc}_{\rm eff} = i \frac{qE}{2 \pi} \int \frac{d^2 {\bf
k}_{\perp}}{(2 \pi)^2} \Bigl[ \ln \sqrt{2 \pi} - \ln \Gamma (-p^*) -
i \frac{(p^*+1)\pi}{2} \Bigr]. \label{eff-E2}
\end{eqnarray}
Here, $qE/(2 \pi)$ is the number of states along the $z$-direction
and $d^2 {\bf k}_{\perp}/(2 \pi)^2$ is the number of states for each
${\bf k}_{\perp}$. Using the gamma function \cite{GR3}
\begin{eqnarray}
\ln \Gamma (z) = \int_0^{\infty} \Bigl[\frac{e^{-z s}}{1 - e^{-s}} -
\frac{e^{-s}}{1 - e^{-s}} + (z-1) e^{-s} \Bigr] \frac{ds}{s},
\label{gamma}
\end{eqnarray}
and subtracting all divergent terms which are independent of $qE$
and linear in $qE$, the effective action per unit volume takes the
form
\begin{eqnarray}
{\cal L}^{\rm sc}_{\rm eff} = -i \frac{qE}{4 \pi} \int \frac{d^2
{\bf k}_{\perp}}{(2 \pi)^2} \int_0^{\infty} \frac{ds}{s}
e^{(p^{*}+1/2)s} \Bigl[ \frac{1}{\sinh(s/2)}
-\frac{2}{s}+\frac{s}{12} \Bigr]. \label{eff-E3}
\end{eqnarray}
Note that the effective action (\ref{eff-E3}) is now finite after a
renormalization prescription, where the $(2/s)$-subtraction,
independent of $qE$, corresponds to the vacuum-energy
renormalization, and the $(-s/12)$-subtraction, quadratic in $qE$,
corresponds to the charge renormalization. Also, note that the
resonances at $p^* = n$ of scattering amplitude, $1/\mu^*_{\bf k}$,
which correspond to the complex energy eigenvalues, $m^2 + {\bf
k}^2_{\perp} + i qE (2n+1)$, from the electric-magnetic duality in
Ref. \cite{Dunne-Hall}, are regular points of the integral in Eq.
(\ref{eff-E3}).

Further, by integrating over the momentum ${\bf k}_{\perp}$, the
effective action is given by a finite integral
\begin{eqnarray}
{\cal L}^{\rm sc}_{\rm eff} &=& \frac{1}{16 \pi^2} \int_0^{\infty}
\frac{ds}{s^3} e^{ im^2 s} \Bigl[ \frac{qE s}{\sinh(qE s)} -1
+\frac{(eE s)^{2}}{6} \Bigr]. \label{eff-E4}
\end{eqnarray}
\begin{figure}[htp]
\begin{center}
\includegraphics[height=2.5in,width=2.5in]{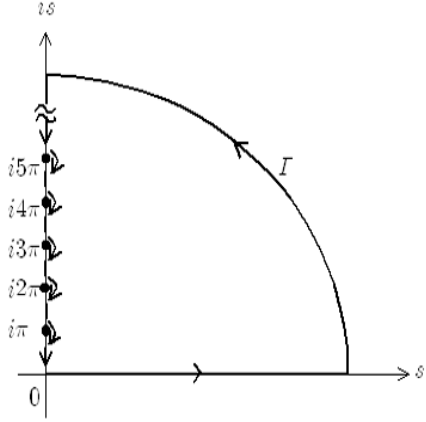}
\caption{The contour of $s$ integration for Eq. (\ref{eff-E5})}
\end{center}
\end{figure}
Finally, by doing the contour integral over a quarter circle in the
first quadrant in Fig. 1, we obtain the exact one-loop effective
action per unit volume
\begin{eqnarray}
{\cal L}^{\rm sc}_{\rm eff} &=& - \frac{1}{16 \pi^2} {\cal P}
\int_0^{\infty} \frac{ds}{s^3} e^{ -m^2 s} \Bigl[ \frac{qE
s}{\sin(qE s)} -1 -\frac{(eE s)^{2}}{6} \Bigr] + i
\frac{(qE)^{2}}{16\pi^{3}} \sum_{n=1}^{\infty}
\frac{(-1)^{n+1}}{n^{2}} e^{-\frac{\pi m^{2}n}{qE}},
\label{scalar-eff}
\end{eqnarray}
where ${\cal P}$ denotes the principal value. The effective action
(\ref{scalar-eff}) agrees with the exact result by Schwinger
\cite{Schwinger}. The imaginary part may be written in the form
\begin{eqnarray}
{\rm Im} ({\cal L}^{\rm sc}_{\rm eff}) = \frac{qE}{2 (2\pi)} \int
\frac{d^2 {\bf k}_{\perp}}{(2\pi)^2} \ln (1 + {\cal N}_{\bf
k}),
\end{eqnarray}
where ${\cal N}_{\bf k} = e^{- \pi (m^2 + {\bf
k}_{\perp}^2)/(qE)}$, and confirms the relation (\ref{scalar-rel}),
now renormalized one.

A passing remark is that the Bogoliubov coefficients (\ref{E-bcoef})
can also be obtained from Eq. (\ref{sol-bcoef}) by appropriately
normalizing the ingoing wave function as
\begin{eqnarray}
\varphi_{{\bf k}, {\rm in}} (t) = \frac{e^{- \pi \frac{m^2 + {\bf
k}_{\perp}^2}{4 qE}}}{\sqrt{2 qE}} D_p (z),
\end{eqnarray}
and by choosing $t_{\rm in} + t_{\rm out} = 2 k_z/qE$, which amounts
to removing the dynamical phase.

\subsection{Spinor QED}

In spinor QED, the effective action per unit volume can be obtained
from Eq. (\ref{spinor-E}) as
\begin{eqnarray}
{\cal L}^{\rm sp}_{\rm eff} = - i \frac{qE}{2 \pi} \int \frac{d^2
{\bf k}_{\perp}}{(2 \pi)^2} \Bigl[ \ln \sqrt{2 \pi} - \ln \Gamma
(-p^*) - i \frac{(p^*+1)\pi}{2} \Bigr]. \label{spin-E1}
\end{eqnarray}
Using the gamma function (\ref{gamma}) and summing over the spin
states $(\sigma = \pm 1/2)$, we find the effective action per unit
volume in the form
\begin{eqnarray}
{\cal L}^{\rm sp}_{\rm eff} = -i \frac{qE}{4 \pi} \int \frac{d^2
{\bf k}_{\perp}}{(2 \pi)^2} \int_0^{\infty} \frac{ds}{s} e^{i
\frac{m^2 + {\bf k}_{\perp}^2 }{2 qE} s} \Bigl[ \coth(s/2)
-\frac{2}{s}+\frac{s}{12} \Bigr]. \label{spin-E2}
\end{eqnarray}
Finally, integrating over the momentum ${\bf k}_{\perp}$ and doing
the contour integral as in the case of scalar QED, we obtain the
renormalized effective action
\begin{eqnarray}
{\cal L}^{\rm sp}_{\rm eff} &=& \frac{1}{8 \pi^2} {\cal P}
\int_0^{\infty} \frac{ds}{s^3} e^{ - m^2 s} \Bigl[ (qEs) \cot(qE s)
- \frac{1}{s} + \frac{(eE s)^{3}}{3} \Bigr] + i \frac{(qE)^{2}}{8
\pi^{3}} \sum_{n=1}^{\infty} \frac{1}{n^{2}} e^{-\frac{\pi
m^{2}n}{qE}}, \label{eff-E5}. \label{spin-E3}
\end{eqnarray}
The effective action (\ref{spin-E3}) agrees with the exact result by
Schwinger \cite{Schwinger}. The imaginary part may be written in the
form
\begin{eqnarray}
{\rm Im} ({\cal L}^{\rm sp}_{\rm eff}) = - \frac{qE}{(2\pi)} \int \frac{d^2
{\bf k}_{\perp}}{(2\pi)^2} \ln (1 - {\cal N}_{\bf k}),
\end{eqnarray}
where ${\cal N}_{\bf k} = e^{- \pi (m^2 + {\bf k}_{\perp}^2)/(qE)}$,
and confirms the relation (\ref{spin-rel}), which is now renormalized.

\section{Effective Action in a Sauter-type Electric Field}

In this section we turn to the main result of this paper, the exact
one-loop effective action for a Sauter-type electric field $E(t) =
E_0 {\rm sech}^2 (t/\tau)$ \cite{Sauter32} together with or without
a parallel constant magnetic field. In a Lorentz frame where the
electric field is parallel to the magnetic field, we may choose as
\begin{eqnarray}
A_{\mu} = \Bigl(0, - \frac{By}{2}, \frac{Bx}{2}, - E_0 \tau (1+
\tanh (\frac{t}{\tau})) \Bigr).
\end{eqnarray}
Here, we have chosen the gauge such that the Klein-Gordon or Dirac
equation reduces to a free scalar theory in the past infinity ($t =
- \infty$) when $B = 0$. In the first case of a pure electric field
($B = 0$),  the time component of the Klein-Gordon equation or the
spin diagonal component of the Dirac equation may be written as
\begin{equation}
\ddot{\varphi}_{\bf k} (t) + \omega_{\bf k}^2 (t) \varphi_{\bf k}
(t) = 0, \label{aux eq}
\end{equation}
where
\begin{eqnarray}
\omega_{\bf k}^2 (t) = \Bigl(k_z - qE_0 \tau (1 + \tanh
(\frac{t}{\tau})) \Bigr)^2 + {\bf k}^2_{\perp} +m^2 + 2 i  \sigma q
E(t),
\end{eqnarray}
where $\sigma = \pm 1/2$ denotes the spin state. The mode equation
(\ref{aux eq}) has the asymptotic frequency at $t = - \infty$ and $+
\infty$,
\begin{eqnarray}
\omega_{{\bf k}, {\rm in}} = \sqrt{{\bf k}^2 +m^2}, \quad
\omega_{{\bf k}, {\rm out}} = \sqrt{(k_z - 2 qE_0 \tau)^2 + {\bf
k}^2_{\perp} +m^2}.
\end{eqnarray}

\subsection{Scalar QED}

The Bogoliubov coefficients between the ingoing and the outgoing
vacua are found from the solution to Eq. (\ref{aux eq}) with $\sigma
= 0$, which has the asymptotic ingoing solution at $t = - \infty$,
\begin{eqnarray}
\varphi_{{\bf k}, {\rm in}} (t) = \frac{ e^{- i \omega_{{\bf k},
{\rm in}} t}}{\sqrt{2 \omega_{{\bf k}, {\rm in}}}}. \label{sol-in}
\end{eqnarray}
The solution is given by
\cite{Narozhnyi-Nikishov,AHN,Kim-Page02,Kim-Lee07}
\begin{eqnarray}
\varphi_{\bf k} (t) = \frac{1}{\sqrt{2 \omega_{{\bf k}, {\rm in}}
e^{\pi \tau \omega_{{\bf k}, {\rm in}}}}} (1 - z)^{1/2 + i
\lambda^{\rm sc}} z^{- i \tau \omega_{{\bf k}, {\rm in}}/2}
F(\alpha_{\bf k}, \beta_{\bf k}; \gamma_{\bf k}; z), \label{sol}
\end{eqnarray}
where $F(\alpha_{\bf k}, \beta_{\bf k}; \gamma_{\bf k}; z)$ is the
hypergeometric function, and
\begin{eqnarray}
z = - e^{2 t/\tau}, \quad \lambda^{\rm sc} = \sqrt{(qE_0 \tau^2)^2 -
\frac{1}{4}},
\end{eqnarray}
and
\begin{eqnarray}
\alpha_{\bf k} &=& \frac{1}{2} - \frac{i}{2} \bigl(\tau \omega_{{\bf
k}, {\rm in}}
- \tau \omega_{{\bf k}, {\rm out}}  - 2 \lambda^{\rm sc} \bigr), \nonumber \\
\beta_{\bf k} &=& \frac{1}{2} - \frac{i}{2} \bigl(\tau \omega_{{\bf
k}, {\rm in}}
+ \tau \omega_{{\bf k}, {\rm out}}  - 2 \lambda^{\rm sc} \bigr), \nonumber \\
\gamma_{\bf k} &=& 1 - i \tau \omega_{{\bf k}, {\rm in}}.
\end{eqnarray}
The solution evolves to the form at $t = \infty$
\begin{eqnarray}
\varphi_{\bf k} (t) = \mu_{\bf k} \varphi_{{\bf k}, {\rm out}} (t) +
\nu_{\bf k} \varphi^*_{{\bf k}, {\rm out}} (t),  \label{asym fr+}
\end{eqnarray}
where
\begin{eqnarray}
\varphi_{{\bf k}, {\rm out}} (t) = \frac{e^{- i \omega_{{\bf k},
{\rm out}} t}}{\sqrt{2 \omega_{{\bf k}, {\rm out}}}} \label{asym
out}
\end{eqnarray}
is the outgoing solution, and
\begin{eqnarray}
\mu_{\bf k} &=& 2^{- i \tau \omega_{{\bf k}, {\rm out}}}\sqrt{
\frac{ \omega_{{\bf k}, {\rm out}}}{\omega_{{\bf k}, {\rm in}}}}
\Biggl( \frac{ \Gamma (\gamma_{\bf k}) \Gamma(\beta_{\bf k} -
\alpha_{\bf k})}{ \Gamma (\beta_{\bf k})
\Gamma (\gamma_{\bf k} - \alpha_{\bf k})} \Biggr), \nonumber \\
\nu_{\bf k} &=& 2^{i \tau \omega^{\rm out}_{\bf k}}
\sqrt{\frac{\omega_{{\bf k}, {\rm out}}}{\omega_{{\bf k}, {\rm
in}}}} \Biggl( \frac{ \Gamma (\gamma_{\bf k}) \Gamma(\alpha_{\bf k}
- \beta_{\bf k})}{\Gamma (\alpha_{\bf k}) \Gamma (\gamma_{\bf k} -
\beta_{\bf k})} \Biggr). \label{mu-inE}
\end{eqnarray}

The terms in front of the brackets of the Bogoliubov coefficient
(\ref{mu-inE}) are removed in a renormalization procedure; thus,
according to Eq. (\ref{eff-E}), the effective action per unit volume
is given by
\begin{eqnarray}
{\cal L}^{\rm sc}_{\rm eff} = i \int \frac{d^3 {\bf k}}{(2 \pi)^3}
\Bigl[ \ln \Gamma (\gamma_{\bf k}^*) + \ln \Gamma(\beta_{\bf k}^* -
\alpha_{\bf k}^*) - \ln \Gamma (\beta_{\bf k}^*) - \ln
\Gamma(\gamma_{\bf k}^* - \alpha_{\bf k}^*) \Bigr]. \label{eff-inE}
\end{eqnarray}
By using the gamma function (\ref{gamma}), we obtain the effective
action in an intermediate form
\begin{eqnarray}
{\cal L}^{\rm sc}_{\rm eff} &=& - \frac{i}{2} \int \frac{d^3 {\bf
k}}{(2 \pi)^3} \int_0^{\infty} \frac{ds}{s} ( e^{-\frac{i}{2}
\Omega_{\bf k}^{(+)} s} + e^{-\frac{i}{2} \Omega_{\bf k}^{(-)}s} )
\Bigl[
\frac{1}{\sinh(s/2)} -\frac{2}{s}+\frac{s}{12} \Bigr] \nonumber\\
&& + \frac{i}{2} \int \frac{d^3 {\bf k}}{(2 \pi)^3} \int_0^{\infty}
\frac{ds}{s} e^{-i \tau \omega_{{\bf k}, {\rm in}} s}
\Bigl[\frac{e^{-s/2}}{\sinh(s/2)} -\frac{2}{s} + 1 - \frac{s}{6}
\Bigr] \nonumber\\ && + \frac{i}{2} \int \frac{d^3 {\bf k}}{(2
\pi)^3} \int_0^{\infty} \frac{ds}{s} e^{-i \tau \omega_{{\bf k},
{\rm out}} s} \Bigl[\frac{e^{s/2}}{\sinh(s/2)} - \frac{2}{s} - 1 -
\frac{s}{6} \Bigr], \label{eff-inE3}
\end{eqnarray}
where
\begin{eqnarray}
\Omega_{\bf k}^{(\pm)} = \tau \omega_{{\bf k}, {\rm in}} + \tau
\omega_{{\bf k}, {\rm out}} \pm 2 \lambda^{\rm sc}.
\end{eqnarray}
Finally, by doing the contour integral over a quarter circle in the
fourth quadrant, we obtain the exact one-loop effective action per
unit volume
\begin{eqnarray}
{\cal L}^{\rm sc}_{\rm eff} &=& \frac{1}{2} \int \frac{d^3 {\bf
k}}{(2 \pi)^3} {\cal P} \int_0^{\infty} \frac{ds}{s} (e^{-
\Omega_{\bf k}^{(+)} s} + e^{- \Omega_{\bf k}^{(-)}s} ) \Bigl[
\frac{1}{\sin(s)} - \frac{1}{s} -
\frac{s}{6} \Bigr] \nonumber\\
&&- \frac{1}{2} \int \frac{d^3 {\bf k}}{(2 \pi)^3} {\cal P}
\int_0^{\infty} \frac{ds}{s} (e^{- 2 \tau \omega_{{\bf k}, {\rm in}}
s} + e^{- 2 \tau \omega_{{\bf k}, {\rm out}}s} ) \Bigl[ \cot (s) -
\frac{1}{s} + \frac{s}{3} \Bigr] \nonumber\\ && + \frac{i}{2} \int
\frac{d^3 {\bf k}}{(2 \pi)^3} \ln \Bigl[ \frac{ ( 1+ e^{- \pi
\Omega_{\bf k}^{(+)}}) (1 + e^{- \pi \Omega_{\bf k}^{(-)}}) }{( 1-
e^{- 2 \pi \tau \omega_{{\bf k}, {\rm in}}}) (1 - e^{- 2 \pi \tau
\omega_{{\bf k}, {\rm out}}})} \Bigr].\label{eff-inE5}
\end{eqnarray}
It can be shown by a direct calculation that
\begin{eqnarray}
2 {\rm Im} ({\cal L}^{\rm sc}_{\rm eff}) = \int \frac{d^3 {\bf k}
}{(2 \pi)^3} \ln (1 + {\cal N}^{\rm sc}_{\bf k}), \label{im-ex rel}
\end{eqnarray}
where
\begin{eqnarray}
{\cal N}^{\rm sc}_{\bf k} = \frac{\cosh (2 \pi \lambda^{\rm sc}) +
\cosh(\pi \tau \omega_{{\bf k}, {\rm out}} - \pi \tau \omega_{{\bf
k}, {\rm in}})}{2 \sinh (\pi \tau \omega_{{\bf k}, {\rm in}})
\sinh(\pi \tau \omega_{{\bf k}, {\rm out}})}.
\label{sauter-scalar-mean}
\end{eqnarray}
The mean number of created pairs, Eq. (\ref{sauter-scalar-mean}),
agrees with the exact result, Eq. (43), of Ref.
\cite{Gavrilov-Gitman}.

In the second case of the electric field together with a constant
magnetic field, the mode equation takes the form
\begin{eqnarray}
\ddot{\varphi}_{k} (t, x, y) + \Biggl[ - (\partial_x^2 +
\partial_y^2) + \Bigl(\frac{qB}{2} \Bigr)^2 ( x^2 + y^2) +
qB L_z + \Bigl(k_z - qE_0 \tau (1+ \tanh (\frac{t}{\tau})) \Bigr)^2
+ m^2 \Biggr] \varphi_{k} (t, x, y) = 0. \label{EB eq}
\end{eqnarray}
Now, the mode equation (\ref{EB eq}) has the asymptotic frequency,
\begin{eqnarray}
\omega_{nk, {\rm in}} = \sqrt{k_z^2 + qB(2n+1) +m^2}, \quad
\omega_{nk, {\rm out}} = \sqrt{(k_z - 2 qE_0 \tau)^2 + qB(2n+1)
+m^2}.
\end{eqnarray}
In Ref. \cite{Kim-Lee07}, the solution is found to be $\varphi_{k}
(t, x, y)= \varphi_{nk} (t) \varphi_{n}(x,y)$, where
$\varphi_{n}(x,y)$ is the two-dimensional harmonic oscillator
function with the Landau level, $qB(2n+1)$, and
\begin{eqnarray}
\varphi_{nk} (t) = \frac{1}{\sqrt{2 \omega_{nk, {\rm in}} e^{\pi
\tau \omega_{nk, {\rm in}}}}} (1 - z)^{1/2 + i \lambda^{\rm sc}}
z^{- i \tau \omega_{nk, {\rm in}}/2} F(\alpha_{nk}, \beta_{nk};
\gamma_{nk}; z),
\end{eqnarray}
with parameters replaced by
\begin{eqnarray}
\alpha_{nk} &=& \frac{1}{2} - \frac{i}{2} \bigl(\tau \omega_{nk,
{\rm in}}
- \tau \omega_{nk, {\rm out}}  - 2 \lambda^{\rm sc} \bigr), \nonumber\\
\beta_{nk} &=& \frac{1}{2} - \frac{i}{2} \bigl(\tau \omega_{nk, {\rm
in}}
+ \tau \omega_{nk, {\rm out}}  - 2 \lambda^{\rm sc} \bigr), \nonumber \\
\gamma_{nk} &=& 1 - i \tau \omega_{nk, {\rm in}}.
\end{eqnarray}
By repeating the same procedure for the pure electric field, we
obtain the exact one-loop effective action per unit volume
\begin{eqnarray}
{\cal L}^{\rm sc}_{\rm eff} &=& \frac{1}{2} \int \frac{d k_z}{2 \pi}
{\cal P} \int_0^{\infty} \frac{ds}{s} \sum_{n = 0}^{\infty} (e^{-
\Omega_{n k}^{(+)} s} + e^{- \Omega_{nk}^{(-)}s} ) \Bigl[
\frac{1}{\sin(s)} - \frac{1}{s} +
\frac{s}{6} \Bigr] \nonumber\\
&&- \frac{1}{2} \int \frac{d k_z}{2 \pi} {\cal P} \int_0^{\infty}
\frac{ds}{s} \sum_{n = 0}^{\infty} (e^{- 2 \tau \omega_{{n k}, {\rm
in}} s} + e^{- 2 \tau \omega_{{nk}, {\rm out}} s} ) \Bigl[ \cot (s)
- \frac{1}{s} + \frac{s}{3} \Bigr] \nonumber\\ &&+ \frac{i}{2}\int
\frac{d k_z}{2 \pi} \sum_{n = 0}^{\infty} \ln \Bigl[ \frac{ (1+ e^{-
\pi \Omega_{n k}^{(+)}}) (1 + e^{- \pi \Omega_{n k}^{(-)}})}{(1 -
e^{- 2 \pi \tau \omega_{{n k}, {\rm out}}})(1- e^{- 2 \pi \tau
\omega_{{n k}, {\rm in}}}) }\Bigr].\label{eff-inE6}
\end{eqnarray}
where
\begin{eqnarray}
\Omega_{n k}^{(\pm)} = \tau \omega_{nk, {\rm in}} + \tau \omega_{nk,
{\rm out}} \pm 2 \lambda^{\rm sc}.
\end{eqnarray}
It can be shown that $2 \, {\rm Im} ({\cal L}^{\rm sc}_{\rm eff}) =
\sum_{\bf k} \ln ( 1+ {\cal N}^{\rm sc}_{\bf k})$, thus confirming
the renormalized relation of Eq. (\ref{scalar-rel}).

\subsection{Spinor QED}

The Sauter-type electric field in spinor QED does not change the
asymptotic solutions $\varphi_{{\bf k}, {\rm in}}$ and
$\varphi_{{\bf k}, {\rm out}}$ for each spin diagonal component,
which define the ingoing and the outgoing vacua. The spin diagonal
component equation (\ref{aux eq}) has the solution
\cite{Narozhnyi-Nikishov,AHN}
\begin{eqnarray}
\varphi_{\bf k} (t) = \frac{1}{\sqrt{2 \omega_{{\bf k}, {\rm in}}
e^{\pi \tau \omega_{{\bf k}, {\rm in}}}}} (1 - z)^{(1- 2 \sigma)/2 +
i \lambda^{\rm sp}} z^{- i \tau \omega_{{\bf k}, {\rm in}}/2}
F(\alpha^{\rm sp}_{\bf k}, \beta^{\rm sp}_{\bf k}; \gamma_{\bf k};
z), \label{sp-sol}
\end{eqnarray}
where
\begin{eqnarray}
z = - e^{2 t/\tau}, \quad \lambda^{\rm sp} = qE_0 \tau^2,
\end{eqnarray}
and
\begin{eqnarray}
\alpha^{\rm sp}_{\bf k} &=& \frac{1- 2 \sigma}{2} - \frac{i}{2}
\bigl(\tau \omega_{{\bf k}, {\rm in}}
- \tau \omega_{{\bf k}, {\rm out}}  - 2 \lambda^{\rm sp} \bigr), \nonumber \\
\beta^{\rm sp}_{\bf k} &=& \frac{1- 2 \sigma}{2} - \frac{i}{2}
\bigl(\tau \omega_{{\bf k}, {\rm in}}
+ \tau \omega_{{\bf k}, {\rm out}}  - 2 \lambda^{\rm sp} \bigr), \nonumber \\
\gamma_{\bf k} &=& 1 - i \tau \omega_{{\bf k}, {\rm in}}.
\end{eqnarray}
Then the Bogoliubov coefficients are
\begin{eqnarray}
\mu^{\rm sp}_{\bf k} &=& 2^{- i \tau \omega_{{\bf k}, {\rm
out}}}\sqrt{ \frac{ \omega_{{\bf k}, {\rm out}}}{\omega_{{\bf k},
{\rm in}}}} \Biggl( \frac{ \Gamma (\gamma_{\bf k}) \Gamma(\beta^{\rm
sp}_{\bf k} - \alpha^{\rm sp}_{\bf k})}{ \Gamma (\beta^{\rm sp}_{\bf
k})
\Gamma (\gamma_{\bf k} - \alpha^{\rm sp}_{\bf k})} \Biggr), \nonumber \\
\nu^{\rm sp}_{\bf k} &=& 2^{i \tau \omega^{\rm out}_{\bf k}}
\sqrt{\frac{\omega_{{\bf k}, {\rm out}}}{\omega_{{\bf k}, {\rm
in}}}} \Biggl( \frac{ \Gamma (\gamma_{\bf k}) \Gamma(\alpha^{\rm
sp}_{\bf k} - \beta^{\rm sp}_{\bf k})}{\Gamma (\alpha^{\rm sp}_{\bf
k}) \Gamma (\gamma_{\bf k} - \beta^{\rm sp}_{\bf k})} \Biggr),
\label{mu-spin-inE}
\end{eqnarray}
and the effective action per unit volume is given by
\begin{eqnarray}
{\cal L}^{\rm sp}_{\rm eff} = i \int \frac{d^3 {\bf k}}{(2 \pi)^3}
\Bigl[ \ln \Gamma (\gamma_{\bf k}^*) + \ln \Gamma(\beta^{\rm sp
*}_{\bf k} - \alpha^{\rm sp *}_{\bf k}) - \ln \Gamma (\beta^{\rm
sp *}_{\bf k}) - \ln \Gamma(\gamma_{\bf k} - \alpha^{\rm sp *}_{\bf
k}) \Bigr]. \label{eff-inE}
\end{eqnarray}
Finally, using the gamma function (\ref{gamma}) and summing over the
spin states, we obtain the exact one-loop effective action in spinor
QED
\begin{eqnarray}
{\cal L}^{\rm sp}_{\rm eff} &=& - \int \frac{d^3 {\bf k}}{(2 \pi)^3}
{\cal P} \int_0^{\infty} \frac{ds}{s} \Bigl[ (e^{- \Omega^{\rm sp
(+)}_{\bf k} s} + e^{- \Omega^{\rm sp (-)}_{\bf k} s}) - (e^{- 2
\tau \omega_{{\bf k}, {\rm in}} s} + e^{- 2 \tau \omega_{{\bf k},
{\rm out}}s} ) \Bigr] \Bigl( \cot (s) - \frac{1}{s} + \frac{s}{3}
\Bigr)
\nonumber\\
&& - i \int \frac{d^3 {\bf k}}{(2 \pi)^3} \ln \Bigl[ \frac{( 1 -
e^{- \pi \Omega^{\rm sp (+)}_{\bf k}}) (1 -  e^{- \pi \Omega^{\rm sp
(-)}_{\bf k}})}{( 1- e^{- 2 \pi \tau \omega_{{\bf k}, {\rm in}}}) (1
- e^{- 2 \pi \tau \omega_{{\bf k}, {\rm out}}})} \Bigr],
\label{spin-sauter-eff}
\end{eqnarray}
where
\begin{eqnarray}
\Omega_{\bf k}^{\rm sp (\pm)} = \tau \omega_{{\bf k}, {\rm in}} +
\tau \omega_{{\bf k}, {\rm out}} \pm 2 \lambda^{\rm sp}.
\end{eqnarray}
Note that the effective action (\ref{spin-sauter-eff}) is finite due
to the renormalization of vacuum-energy and charge. The difference
of the factor of two from Eq. (\ref{eff-inE5}) in scalar QED is the
spin multiplicity of spin-1/2 fermions. A direct calculation leads
to the general relation, which is renormalized,
\begin{eqnarray}
{\rm Im} ({\cal L}^{\rm sp}_{\rm eff}) = - \int \frac{d^3 {\bf
k}}{(2 \pi)^3} \ln (1 - {\cal N}^{\rm sp}_{\bf k}),
\end{eqnarray}
where
\begin{eqnarray}
{\cal N}^{\rm sp}_{\bf k} = \frac{\cosh (2 \pi \lambda^{\rm sp}) -
\cosh(\pi \tau \omega_{{\bf k}, {\rm out}} - \pi \tau \omega_{{\bf
k}, {\rm in}})}{2 \sinh (\pi \tau \omega_{{\bf k}, {\rm in}})
\sinh(\pi \tau \omega_{{\bf k}, {\rm out}})}.
\label{sauter-spin-mean}
\end{eqnarray}

A few comments are in order. First, the mean number
(\ref{sauter-spin-mean}) from the imaginary part agrees with the
exact result (42) for spin-1/2 fermions of Ref.
\cite{Gavrilov-Gitman} and also with Ref. \cite{Narozhnyi-Nikishov}.
Second, we may compare the imaginary part with Eq. (30) of Ref.
\cite{Dunne-Hall},
\begin{eqnarray}
{\rm Im} ({\cal L}^{\rm sp}_{\rm eff}) = \int \frac{d^3 {\bf k} }{(2
\pi)^3} \ln \Bigl[ ( 1 - e^{- \pi \Omega_{\bf k}^{(+)}}) (1 - e^{-
\pi \Omega_{\bf k}^{(-)}}) \Bigr]. \label{sp im-inE}
\end{eqnarray}
However, the denominator of the imaginary part in Eq.
(\ref{spin-sauter-eff}) is necessary to explain no pair
production in the limit of either $E_0 =0$ for a finite $\tau$ or
$\tau =0$ for a finite $E_0$. Note that in
the limit of $\tau \omega_{{\bf k}, {\rm in}}
\gg 1$ the denominator
approaches to identity and the imaginary part of Eq.
(\ref{spin-sauter-eff}) reduces to Eq. (\ref{sp im-inE}).

In the presence of a constant magnetic field parallel to the
Sauter-type electric field, the exact one-loop effective action is
obtained by summing over Landau levels,
\begin{eqnarray}
{\cal L}^{\rm sp}_{\rm eff} &=& - \int \frac{d k_z}{2 \pi}  {\cal P}
\int_0^{\infty} \frac{ds}{s} \sum_{n = 0}^{\infty}  \Bigl[ (e^{-
\Omega^{\rm sp (+)}_{n k} s} + e^{- \Omega^{\rm sp (-)}_{n k} s}) -
(e^{- 2 \tau \omega_{{n k}, {\rm in}} s} + e^{- 2 \tau \omega_{{n
k}, {\rm out}}s} ) \Bigr] \Bigl( \cot (s) - \frac{1}{s} +
\frac{s}{3} \Bigr)
\nonumber\\
&& - i \int \frac{d k_z}{2 \pi}  {\cal P} \int_0^{\infty}
\frac{ds}{s} \sum_{n = 0}^{\infty} \ln \Bigl[ \frac{( 1 - e^{- \pi
\Omega^{\rm sp (+)}_{n k}}) (1 -  e^{- \pi \Omega^{\rm sp (-)}_{n
k}})}{( 1- e^{- 2 \pi \tau \omega_{{nk}, {\rm in}}}) (1 - e^{- 2 \pi
\tau \omega_{{n k}, {\rm out}}})} \Bigr], \label{spin-pair-prod}
\end{eqnarray}
where
\begin{eqnarray}
\omega_{nk, {\rm in}} = \sqrt{k_z^2 + qB(2n+1) +m^2}, \quad
\omega_{nk, {\rm out}} = \sqrt{(k_z - 2 qE_0 \tau)^2 + qB(2n+1)
+m^2}.
\end{eqnarray}
and
\begin{eqnarray}
\Omega_{n k}^{\rm sp (\pm)} = \tau \omega_{nk, {\rm in}} + \tau
\omega_{nk, {\rm out}} \pm 2 \lambda^{\rm sp}.
\end{eqnarray}

\section{Conclusion}

In this paper we further developed the evolution operator method to
calculate the effective action of scalar and spinor QED in electric
field backgrounds, which was expressed in terms of the Bogoliubov
coefficient. In fact, the Bogoliubov transformation between the
ingoing and the outgoing Fock space enables one to explicitly
calculate the scattering amplitude between the ingoing and the
outgoing vacua, whose complex phase is the effective action. For
that purpose, we first expressed the evolution operator in terms of
a two-mode squeeze operator and an overall phase part. As the
two-mode squeeze operator carries the information about the
multi-pair production, this approach correctly yields not only the
mean number of created pairs but also the vacuum polarization.

We first applied the method to a constant electric field to clarify
the renormalization procedure. The exact one-loop effective action
and its imaginary part thus obtained agreed with the standard scalar
and spinor QED result from other methods. According to the
electric-magnetic duality, the Landau levels of a charged scalar
particle or spin-1/2 fermion in a magnetic field correspond to a
discrete spectrum of complex frequencies in the electric field.
However, the actual motion in an electric field is an acceleration.
Instead, the evolution operator method makes use of the Bogoliubov
transformation between the ingoing and the outgoing Fock space. The
Bogoliubov coefficient leads to the exact one-loop effective action
simultaneously with the correct imaginary part as the sum of
residues of a proper integral.

As the next application, we applied the method to a Sauter-type
electric field, effectively acting for a finite period of time.
Using the Bogoliubov coefficient, we obtained the exact one-loop
effective action at zero temperature, which is the main result of
this paper. The real part of the effective action of scalar and
spinor QED takes the form
\begin{eqnarray}
{\rm Re} ({\cal L}_{\rm eff}) &=& \pm \frac{2S+1}{2} \int \frac{d^3
{\bf k}}{(2 \pi)^3} {\cal P} \int_0^{\infty} \frac{ds}{s} \Bigl[
(e^{- \Omega_{\bf k}^{(+)} s} + e^{- \Omega_{\bf k}^{(-)}s}) f(s) -
(e^{- 2 \tau \omega_{{\bf k}, {\rm in}} s} + e^{- 2 \tau
\omega_{{\bf k}, {\rm out}}s} ) g(s) \Bigr],  \label{vac-pol}
\end{eqnarray}
where the upper (lower) sign and $S=0 ~(1/2)$ are for scalar
particles (spin-1/2 fermions), and $g(s) = \cot (s) - 1/s + s/3$,
and $f(s) = 1/\sin(s) - 1/s + s/6$ for scalar particles and $f(s) =
g(s)$ for spin-1/2 fermions. The imaginary part is given by
\begin{eqnarray}
{\rm Im} ({\cal L}_{\rm eff}) = \pm \frac{2S+1}{2} \int \frac{d^3
{\bf k}}{(2 \pi)^3} \ln \Bigl[ \frac{( 1 \pm e^{- \pi \Omega_{\bf
k}^{(+)}}) (1 \pm  e^{- \pi \Omega_{\bf k}^{(-)}})}{( 1- e^{- 2 \pi
\tau \omega_{{\bf k}, {\rm in}}}) (1 - e^{- 2 \pi \tau \omega_{{\bf
k}, {\rm out}}})} \Bigr], \label{pair-prod}
\end{eqnarray}
where the upper positive signs are for scalar particles and the
lower negative signs for spin-1/2 fermions. It is further shown that
the imaginary part of the renormalized effective action
indeed satisfies the general relation, as expected,
\begin{eqnarray}
2 ({\rm Im} {\cal L}_{\rm eff}) = \pm (2S+1) \int \frac{d^3 {\bf
k}}{(2 \pi)^3} \ln (1 \pm {\cal N}_{\bf k}),
\end{eqnarray}
where
\begin{eqnarray}
{\cal N}_{\bf k} = \frac{\cosh (2 \pi \lambda) \pm \cosh(\pi \tau
\omega_{{\bf k}, {\rm out}} - \pi \tau \omega_{{\bf k}, {\rm
in}})}{2 \sinh (\pi \tau \omega_{{\bf k}, {\rm in}}) \sinh(\pi \tau
\omega_{{\bf k}, {\rm out}})}. \label{sauter-mean}
\end{eqnarray}
The mean number of created pairs, Eq. (\ref{sauter-mean}), agrees
with the exact results, Eq. (43) for scalar particles and Eq. (42)
for spin-1/2 fermions of Ref. \cite{Gavrilov-Gitman}. The
denominator in Eq. (\ref{pair-prod}), which is missing in Ref.
\cite{Dunne-Hall}, is necessary to explain no pair
production in the limit of either $E_0 =0$ for a finite $\tau$ or
$\tau =0$ for a finite $E_0$. However, in the other limit of $\tau
\rightarrow 0$ for a finite $E_0 \tau$, the mean number approaches
the limiting value ${\cal N}^{\rm sc}_{\bf k} = (\omega_{{\bf k},
{\rm out}} - \omega_{{\bf k}, {\rm in}})^2/(4 \omega_{{\bf k}, {\rm
in}} \omega_{{\bf k}, {\rm out}})$ for scalar particles and ${\cal
N}^{\rm sp}_{\bf k} = [ 4 (qE_0 \tau)^2 - (\omega_{{\bf k}, {\rm
out}} - \omega_{{\bf k}, {\rm in}})^2]/(4 \omega_{{\bf k}, {\rm in}}
\omega_{{\bf k}, {\rm out}})$ for spin-1/2 fermions. Finally, we
found the exact one-loop effective action both in the Sauter-type
electric field and in a constant magnetic field.

It should be pointed out that the method can be applied to the
finite-temperature effective action in a time-dependent electric
field, which describes a nonequilibrium quantum field \cite{KLY}.

\acknowledgments

We would like to thank Prof. S.~P.~Gavrilov for useful comments.
S.~P.~K. appreciates the hospitality of Hanyang University and
Center for Quantum Spacetime (CQUeST) of Sogang University, and
H.~K.~L. and Y.~Y. appreciate the hospitality of Kunsan National
University. The work of S.~P.~K. was supported by the Korea Research
Foundation Grant funded by the Korean Government (MOEHRD)
(KRF-2007-C00167) and the work of H.~K.~L. was supported by the
Korea Science and Engineering Foundation (KOSEF) grant funded by the
Korea government (MOST) (No. R01-2006-000-10651-0).
\appendix

\end{document}